\documentclass{article}

\usepackage{arxiv}

\usepackage[utf8]{inputenc} 
\usepackage[T1]{fontenc}    
\usepackage{hyperref}       
\usepackage{url}            
\usepackage{booktabs}       
\usepackage{amsfonts}       
\usepackage{nicefrac}       
\usepackage{microtype}      
\usepackage{lipsum}
\usepackage{graphicx} 
\title{Open Source Software Development Process: A Systematic Review}

\author{
  Bianca M. Napoleão\\
  Laboratoire d’informatique formelle \\
  Université du Québec à Chicoutimi\\
  \texttt{bianca.minetto-napoleao1@uqac.ca} \\
   \And
  Fabio Petrillo\\
  Laboratoire d’informatique formelle \\
  Université du Québec à Chicoutimi\\
  \texttt{fabio@petrillo.com} \\
    \And
  Sylvain Hallé\\
  Laboratoire d’informatique formelle \\
  Université du Québec à Chicoutimi\\
  \texttt{shalle@acm.org} \\
}

\begin{document}
\maketitle

\begin{abstract}
Open Source Software (OSS) has been recognized by the software development community as an effective way to deliver software.
Unlike traditional software development, OSS development is driven by collaboration among developers spread geographically and motivated by common goals and interests. Besides this fact, it is recognized by OSS community the need of understand OSS development process and its activities.
Our goal is to investigate the state-of-art about OSS process through conducting a systematic literature review providing an overview of how the OSS community has been investigating OSS process over past years identifying and summarizing OSS process activities and their characteristics as well as translating OSS process in a macro process through BPMN notation. 
As a result, we systematically analysed 33 studies presenting an overview of the state-of-art of researches regarding OSS process, a generalized OSS development macro process represented by BPMN notation with a detailed description of each OSS process activity and roles in OSS environment. 
 We conclude that OSS process can be in practice further investigated by researchers.  In addition, the presented OSS process can be used as a guide for OSS projects and being adapted according to each OSS project reality. It provides insights to managers and developers who want to improve their development process even in OSS and traditional environments. Finally, recommendations for OSS community regarding OSS process activities are provided.
\end{abstract}


\section{Introduction}


Nowadays, Open Source Software (OSS) represents an important force in the current scenario of software industrial production \cite{STEINMACHER2018}. According to \cite{BALALI2018}, there are many examples of projects that were developed and maintained through open source initiatives such as  Linux, Open Office, and Mozilla Firefox.  Moreover, companies that have a restrictive business model are publishing products and artifacts via OSS \cite{PINTO2018}. 
Due to the significant attention it has gained over the years, OSS has been widely recognized as an effective way to deliver software \cite{10KRISHNAMURTHY2013, PINTO2018}. The OSS principle is based on continuous interaction in a participatory user community motivated by common goals and interests; it introduces non-conventional ways to develop, test, and maintain software \cite{GGLASSMAN2016, 4EZEALA2008}.  

There are several studies presenting investigations on how the OSS process differs from traditional Software Engineering (SE) processes \cite{2SCACCHI2006, 27POTDAR2004, 33FITZGERALD2005}. According to \cite{2SCACCHI2006}, OSS projects have complex development processes executed by weakly coordinated contributors and developers spread geographically. Also, as stated by \cite{28FUGGETTA2003}, it is difficult to detect aspects of OSS development compared to other traditional software development practices.  As a matter of fact, OSS projects have received criticism for not making their development process clear. Therefore, discovering and mapping the process and its activities is a recognized need by the OSS community \cite{14HASHMI2010, 2SCACCHI2006}. One obstacle to developers starting to collaborate in an OSS is precisely a lack of a formal description of the different activities and phases performed in an OSS project \cite{26SENYARD2004}. 

In this paper, we investigate the current state-of-art regarding OSS process by conducting a Systematic Literature Review (SLR), which provides an overview of how the OSS community has been investigating the OSS process over past years. We identify and summarize OSS process activities and their characteristics, as well as translate the OSS process into a macro process through the BPMN notation. 

Our results provide a valuable overview of the OSS development process and show that this process can be further investigated by researchers.  In addition, the presented OSS macro process can be used as a guide for OSS projects and also being adapted according to the OSS project reality. The mapped process and the description of its activities provide insights to managers and developers who want to improve their development process in OSS and even traditional environments. Finally, we present recommendations for the OSS community regarding OSS process activities.
 

The rest of the paper is organized as follows. Section \ref{SLR}, describes the design of the study.  Section \ref{results} presents our results answering the proposed research questions. Section \ref{DiscussionandRecommendations} discusses our findings. Recommendations regarding OSS development process are provided in Section \ref{Recommendations}. Section \ref{threatstovalidity} presents threats to validity to this study. Finally, Section \ref{Conclusions} concludes our study including directions for future research.


\section{Study Design}
\label{SLR}
 
This section covers the SLR method performed to understand the main contributions of the state-of-art regarding OSS development process. A SLR is a means of identifying, evaluating, and summarizing available research related to a particular research topic \cite{Kitchenham07}. In this context, a SLR was appropriate to summarize existing evidence regarding OSS development process. 

Two others studies investigated the state-of-art on OSS process. Acuña \textit{et al.}\@ \cite{ACUNA2012} performed a systematic mapping identifying what activities OSS process models contain and grouped them according to its focuses (concept exploration, software requirements, design, maintenance and evaluation). They conclude that use the  primary studies is the start point to analyse and propose a OS process model for OSS community. In this paper, we use primary studies evidence to summarize and propose a OSS macro process.
Crowston \textit{et al.}\@ \cite{CROWSTON2012}  presented a quantitative summary of articles categorized into issues pertaining to inputs,  processes (software development and social processes), emergent states and outputs. 
However, this study does not investigate in detail OSS development process activities and their characteristics or even mapped them in an unified process, as we present. 

To conduct this study, we followed the guidelines for performing SLR developed by \cite{Kitchenham07}. The SLR steps were conducted by three researchers: one research worked in the execution of the search strategy and through weekly meetings, all researchers reviewed the work's progress and discussed any disagreements about the decisions during the review.

\subsection{Research Questions}

Our goal is to understand how the OSS community has been investigating the OSS process over past years, identifying and summarizing OSS process activities and their characteristics, as well as translating the OSS process into a macro process using BPMN notation.  We translated our research goal into two Research Questions (RQs).

\textbf{RQ1: How has the OSS community been investigating the OSS process over the past years?}

This question aims to analyze the intensity of scientific interest on OSS process thorough understanding of how the OSS community has been studying and exploring the OSS process. A categorized map from previous publications  is provided.  

\textbf{RQ2: What activities do OSS processes contain? What are their main characteristics?}

This question intends to identify OSS process activities and summarize their main characteristics to understand the processes behind the OSS projects.  A OSS macro process is presented and explained in detail. 

The adopted search strategy includes an automated search and a snowballing search \cite{Wohlin14}. We performed the automated search in order to detect relevant primary studies related to OSS development process. The results from the automated search showed that over the last years the number of primary studies investigating the OSS process has decreased. For this reason, we decided to perform forward and backward snowballing to avoid missing relevant studies. We followed Wohlin's guidelines to perform the snowballing technique \cite{Wohlin14}. 

Regarding the automated search, we developed a search query and we ran a search strategy pilot test as recommended by \cite{Kitchenham07}. The study control group, used to calibrate our search string, was based on included studies from \cite{ACUNA2012} secondary study, since this study aimed to investigate the OSS development process as well. Our final search query is: 

\begin{center} 
\texttt{(("open source" OR  "free source")  AND  ("development process" OR "software process"))}
\end{center}

We chose to run our search query on the most renowned SE digital libraries \cite{Kitchenham15}: \textit{Scopus, Web of Science, IEEE Xplore} and \textit{ACM Digital Library}. Other digital libraries were not chosen because Scopus and Web of Science digital libraries index studies of several international publishers, including Cambridge University Press, IEEE, Springer, Wiley-Blackwell and Elsevier. We executed the search query in three metadata fields: title, abstract and keywords. Also, the search query was adapted to meet specific search criteria (e.g.\ syntax) of each digital library.

The selection criteria are organized into two Inclusion Criteria (IC) and four Exclusion Criteria (EC):

\begin{itemize}
    \item \textbf{IC1:} The study must address OSS  development process or models in open source projects;
    \item \textbf{IC2:} The study must be a peer-reviewed study.
    \item \textbf{EC1:} The study is just published as an abstract; 
    \item \textbf{EC2:} The study is not written in English;
    \item \textbf{EC3:} The study is an older version of another study already considered;
   \item  \textbf{EC4:} The study addresses process information considering hybrid projects instead of only OSS projects. 
\end{itemize}

\begin{figure}[!h]
    \centering
    \includegraphics [height=0.6\textwidth]{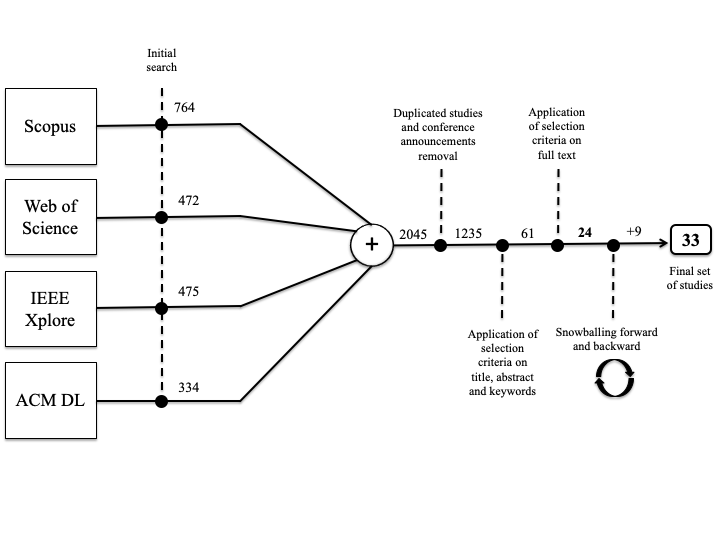}
    \caption{Search strategy process}
    \label{searchprocess}
\end{figure}

As illustrated in Figure \ref{searchprocess}, a total of 2045 items were returned from the automated search execution. Then, we removed all duplicated studies and conference announcements, totaling 1235 studies. Next, we read the papers' title, abstract and keywords and applied the selection criteria (IC and EC) on these fields which reduced our number to 61 candidate studies. Finally, the selection criteria were applied considering the reading of each study's full text, resulting in a set of 24 included studies from this stage. This step was performed by one author, and revised by a second author. 

Regarding the snowballing search, we used the 24 relevant included studies from our automated search process as our seed set (starting set) for performing three iterations. We performed the forward and backward snowballing technique. Forward snowballing is an approach that considers theh analyzed studies' citations,  while backward snowballing considers each study's reference list aiming to find others relevant studies \cite{Wohlin14}.  The studies' citations  were extracted with support of search engines, such as \textit{Google Scholar}, the \textit{ACM Digital Library} and \textit{IEEE Xplore}. In each snowballing iteration, we applied IC and EC criteria first on title, abstract and keywords and next on full text. We performed three snowballing iterations, stopping its execution because in the last iteration no more relevant study was detected. The snowballing technique was performed by one author and when doubts about the inclusion or exclusion of a paper were raised, it was discussed with all authors in the weekly meetings. As a result of the forward snowballing approach, we added nine more relevant studies in our final set of included studies, totaling 33 studies. Figure \ref{searchprocess} illustrates these findings.

\subsection{Data Extraction and Analysis}
In order to extract all relevant data to answer our RQs, we created a data extraction form based on our RQ goals. In other words, the fields of the data extraction form considered the data needed to answer our RQs, for instance, OSS process activities described in the paper, description of each the activity, roles, etc.  The data were extracted by one researcher and reviewed by an experienced researcher who already worked with several OSS projects. As mentioned before, the findings were discussed and reviewed by all author every week. For example, to answer RQ1 and defined the main topic addressed by each paper, all researchers read the papers and though consensus in the meeting the papers' categories were defined.

The data synthesis was performed through a combination of content analysis from the extracted data categorizing the findings into broad thematic categories as well as a narrative synthesis. The extracted information was interpreted and analyzed considering the authors' knowledge and experience in OSS projects. It is worth mentioning that the OSS macro process presented in this paper is a summary of synthesized evidence from primary studies.

\section{Results}
\label{results}

This Section presents our RQs answers based on the results from our included studies data extraction and synthesis. 

\textbf{RQ1: How has the OSS community been investigating the OSS process over the past years?}

\begin{figure}[!h]
    \centering
    \includegraphics [width=0.7\textwidth]{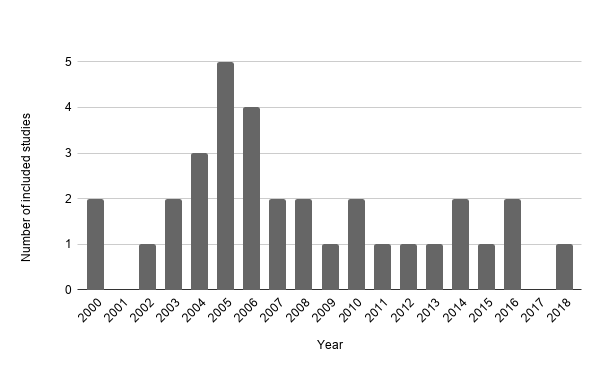}
    \caption{Distribution of publications (included studies) by year}
    \label{publicationyear}
\end{figure}

To analyze the intensity of scientific interest on OSS process, we analyzed the distribution of publications over past years addressed by our included studies. Figure \ref{publicationyear} illustrates our results. 

As shown in Figure \ref{publicationyear}, the highest number of publications regarding the OSS process was published in the years 2004 (three studies), 2005 (five studies) and 2006 (four studies) representing 36.36\% (12 studies) from our set of included studies. There were no included studies published in 2001 and 2017. 
Also in Figure \ref{publicationyear}, it is possible to observe that the interest in OSS process or specific activities about OSS process is still being investigated over the years. For example, continuous integration is a process activity that has been recently adopted in OSS projects \cite{3RHAMAN2018}. 

Through automatic search, the two most recent included studies found were from 2014 and 2016. As mentioned before, we perform a snowballing forward search in order to detect other relevant studies (mainly more recent studies) addressing the OSS process. 

We categorized the included studies according to their main addressed topic. Also, we categorized the publication type (journal, conference, workshop or symposium) of each included study. Table \ref{tab:listofincludedstudies} presents these findings sorted by publication type. 

\setcounter {table}{0}
\begin{table} [!ht]
\centering
\begin{tabular}{l l l}
\hline
\textbf{Included Study} &\textbf{Main topic} &\textbf{Publication type} \\ 
\hline
\hline 
\cite{2SCACCHI2006} & General OSS process & Journal  \\
\hline
\cite{7KUMAR2014} & General OSS process & Journal  \\
\hline
\cite{11SCOTTO2007} & Journal  \\
\hline
\cite{13WANG2015} & Quality assurance & Journal  \\
\hline
\cite{28FUGGETTA2003} & General OSS process & Journal  \\
\hline
\cite{25SCACCHI2007} & General OSS process  & Journal  \\
\hline
\cite{30FITZGERALD2006} & General OSS process & Journal \\
\hline
\cite{24SCACCHI2004}  & General OSS process & Journal  \\
\hline
\cite{18TRONG2005} & General OSS process & Journal  \\
\hline
\cite{19MOCKUS2002} & General OSS process & Journal  \\
\hline
\cite{1MOCKUS2000} & General OSS process  & Conference \\
\hline
\cite{26SENYARD2004} & General OSS process & Conference \\
\hline
\cite{27POTDAR2004} & General OSS process & Conference  \\
\hline
\cite{29DILLON2009}  & General OSS process & Conference   \\
\hline
\cite{31MICHLMAYR2005}  & General OSS process & Conference  \\
\hline
\cite{32SCHROEDER2004} & General OSS process & Conference  \\
\hline
\cite{33FITZGERALD2005} & General OSS process & Conference  \\
\hline
\cite{3RHAMAN2018}  & Continuous integration & Conference \\
\hline
\cite{4EZEALA2008}  & General OSS process & Conference  \\
\hline
\cite{5ABDOU2012}  & Testing & Conference \\
\hline
\cite{15SCACCHI2008} & General OSS process & Conference \\
\hline
\cite{22YAMAUCHI2000} & General OSS process  & Conference  \\
\hline
\cite{17JENSEN2007}  & General OSS process & Conference  \\
\hline
\cite{20ABDOU2015} & Testing & Conference  \\
\hline
\cite{6DECARVALHO2006} & General OSS process & Conference  \\
\hline
\cite{10KRISHNAMURTHY2013}  & General OSS process  & Conference  \\
\hline
\cite{12LAVAZZA2010}  & General OSS process & Conference \\
\hline
\cite{8NEULINGER2016}  & General OSS process & Conference  \\
\hline
\cite{21ROCHA2016}  & Bug & Symposium  \\
\hline
\cite{34KHANJANI2011} & Quality assurance & Symposium  \\
\hline
\cite{23JENSEN2005} & General OSS process & Workshop  \\
 \hline
\cite{9IHARA2009} & Bug  & Workshop  \\
\hline
\cite{14HASHMI2010}  & General OSS process & Workshop  \\
\hline
\end{tabular}
\caption{List of included studies sorted by publication type}
\label{tab:listofincludedstudies}
\end{table}

As Table  \ref{tab:listofincludedstudies} shows, the biggest part of the included studies were published in conferences (17 studies -- 51.51\%) followed by journals (10 studies -- 30.30\%), workshops (3 studies -- 9.09\%) and  symposiums (2 studies -- 6.06\%). Publications from conferences and journals include some of the most renowned SE venues such as the International Conference on Software Engineering, the Journal of Information and Software Technology, the Journal of Systems Software and the Transactions on Software Engineering. 

The majority of the included studies address the OSS process generally. They present information and characteristics regarding the whole OSS process activities and roles. On the other hand, there are studies that address specific activities (sub-processes) or practices in OSS environment, namely testing \cite{5ABDOU2012, 20ABDOU2015}, quality assurance \cite{13WANG2015, 34KHANJANI2011}, bugs \cite{21ROCHA2016,9IHARA2009} and continuous integration \cite{3RHAMAN2018} (see Table \ref{tab:listofincludedstudies}).


\textbf{RQ2: What activities do OSS processes contain? What are their main characteristics?}

OSS development is characterized by collaborative and voluntary software development carried out by several developers distributed geographically around the world \cite{12LAVAZZA2010, 28FUGGETTA2003,25SCACCHI2007}. The distributed aspect in OSS development allows OSS developers to work from the most different locations and even hardly ever or never meeting one each other face-to-face; ordinarily they coordinate and exchange experiences and project information only online (for instance, via emails and bulletin boards) \cite{1MOCKUS2000}. As stated by \cite{1MOCKUS2000, 19MOCKUS2002} the main OSS process characteristics are: (i)~vast number of volunteer developers involved in the project; (ii)~openness for the developer to choose the work they want to do (tasks are not assigned); (iii)~the system under development has no explicit design, the system is built collaboratively; and (iv)~there is no project plan, calendar or even an established delivery list.

To understand the OSS development process phenomenon, firstly we identify and summarize the activities contained in each included study into a macro process using the BPMN notation \cite{BPMNGroup}  as well as provide a detailed explanation of each activity contained in this OSS development macro process. Secondly, we analyze what are the roles played by developers in OSS projects. These results are presented in Section \ref{processactivities},  \ref{processactivitiescaracteristics} and \ref{processroles}, respectively. 

\subsubsection{OSS Process Activities}
\label{processactivities}

\begin{figure*}
    \centering
    \includegraphics [width=1
\textwidth]{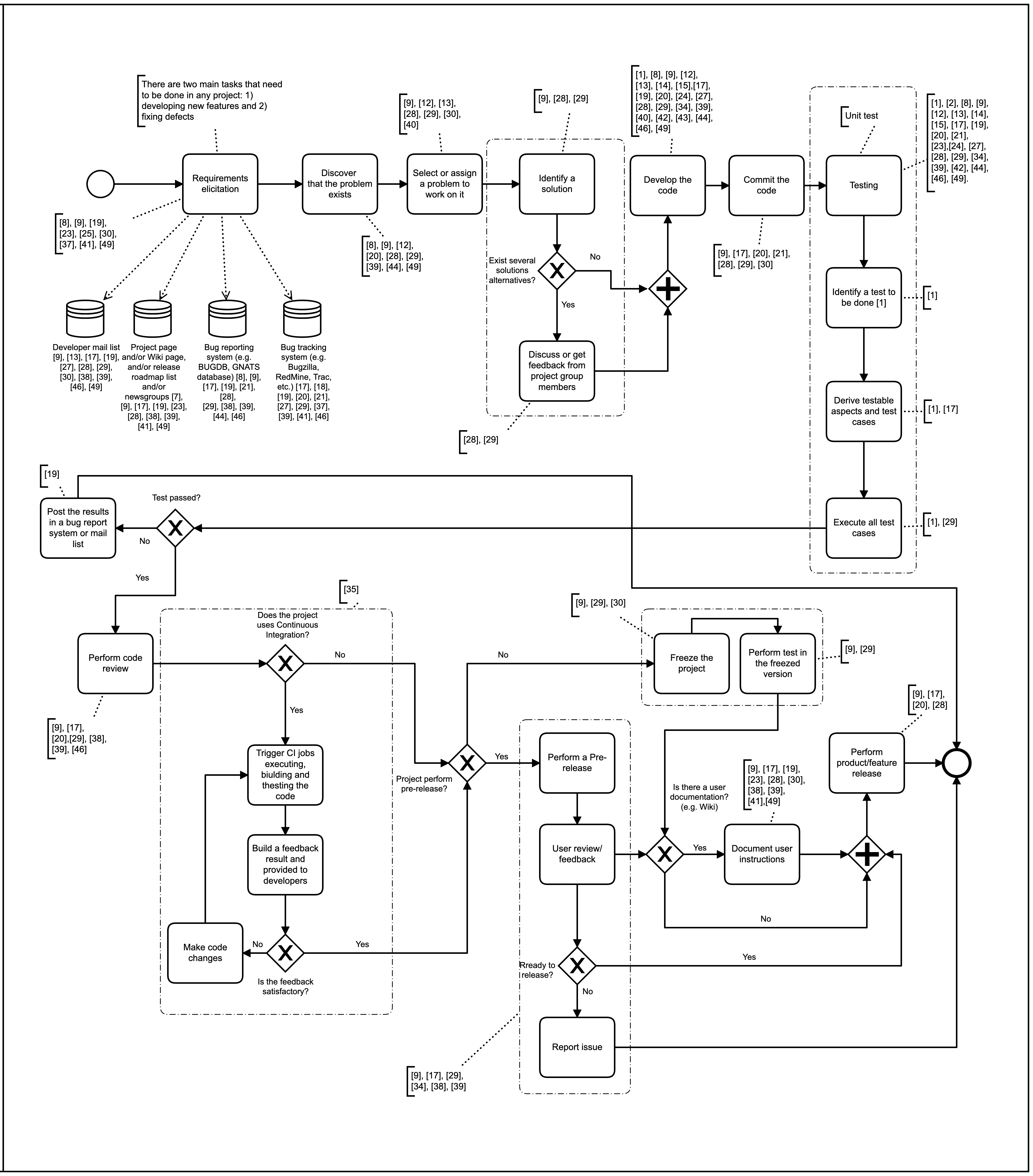}
    \caption{OSS macro process shown in the BPMN notation}
    \label{OSSprocess}
\end{figure*}

We focused on identifying what are the process activities performed by the OSS developers and consolidating evidence into a macro process. Figure \ref{OSSprocess} presents the OSS macro process represented in BPMN notation. BPMN is a standard business process modeling notation that can be applied to modeling software processes \cite{BPMNPilatti2012,BPMNCampos2013}. According to Dumas and Pfahl \cite{BPMNDumas2016}, it offers a valuable spectrum of constructs that can be used to capture software development activities considering process behavioral and structural proprieties in closed and open source environments. 
 
As shown in Figure \ref{OSSprocess}, the OSS process starts with a requirements elicitation step where a developer can report or work on new features or fix defects in an OSS project. These requirements can be even reported or consulted in several sources: developers' mailing lists, project webpage, wiki roadmap or newsgroups and bug reporting or tracking systems. Looking into these sources, a developer discovers that a problem exists or validates if a problem that he/she is facing was reported or not by others. In the sequence, the developer selects a problem to work on it or assigns it as a suggestion to another developer that could contribute to the problem. The following step is to identify a solution to this problem and consequently develop a solution code and commit this code to the project repository. After committing the code, several tests can be performed by the project community. If a test fails, the results should be posted in a mailing list, bug reporting or tracking system (this depending on the bug or issues source used in the development project). On the other hand, if all tests pass, a code review can be performed in order to apply code best practices, improve the code's quality including aspects such as performance, security or consistency.

Some OSS projects adopted a continuous integration practice. Even if this practice is not applied, the next step can be to freeze the project and perform tests on a frozen version or perform a project pre-release to get users' feedback. A pre-release is not a mandatory practice, but it is commonly adopted by several OSS projects. On the one hand, if major issues are found they should be reported. On the other hand, if the project is ready to release, the next step is to write user instructions (if needed) and finally perform a final release. 

\subsubsection{OSS process activities characteristics}
\label{processactivitiescaracteristics}

In the following, we explain in detail each OSS process activity and its characteristics, as illustrated in Figure \ref{OSSprocess}. 

\textbf{Requirements elicitation --} Unlike traditional software processes, requirements elicitation in OSS projects is based on the knowledge and professional experience of core developers \cite{8NEULINGER2016, 15SCACCHI2008}. According to \cite{29DILLON2009}, in OSS development there is no formal requirement process. However, requirements in OSS projects are done basically through two main tasks: (i)~development of new features and (ii)~fixing defects \cite{18TRONG2005, 21ROCHA2016, 22YAMAUCHI2000, 23JENSEN2005}. Requirements in OSS projects emerge from the communication and discussion among OSS community members \cite{12LAVAZZA2010, 8NEULINGER2016}. New features are normally defined by core developers who control the project architecture and direction of development; for example, they look into the project wiki and decide on how and when to start a feature development \cite{10KRISHNAMURTHY2013}. Nonetheless,  in some cases individuals committers can also add a new feature or create a team to work on a large project specifically \cite{18TRONG2005}. It can be said that OSS requirements are community-built requirements \cite{29DILLON2009}.

The main requirement inputs and sources in OSS are developers mail lists \cite{1MOCKUS2000, 18TRONG2005, 13WANG2015, 14HASHMI2010, 19MOCKUS2002, 22YAMAUCHI2000, 23JENSEN2005, 30FITZGERALD2006, 31MICHLMAYR2005, 24SCACCHI2004, 25SCACCHI2007} where developers report issues, give suggestions and discuss bugs \cite{8NEULINGER2016}; project webpage and/or wiki page and/or roadmap list and/or newsgroup \cite{1MOCKUS2000, 6DECARVALHO2006, 10KRISHNAMURTHY2013, 14HASHMI2010, 15SCACCHI2008, 18TRONG2005, 22YAMAUCHI2000, 23JENSEN2005, 24SCACCHI2004, 25SCACCHI2007}  where project information with focus on end users is made available; bug report databases (e.g.\ BUGDB and GNATS) \cite{1MOCKUS2000, 13WANG2015, 14HASHMI2010, 18TRONG2005, 19MOCKUS2002, 23JENSEN2005, 24SCACCHI2004, 25SCACCHI2007, 26SENYARD2004, 29DILLON2009, 34KHANJANI2011} and bug tracking systems (e.g.\ Bugzilla, RedMine and Trac) \cite{9IHARA2009, 13WANG2015, 14HASHMI2010, 15SCACCHI2008, 17JENSEN2007, 19MOCKUS2002, 21ROCHA2016, 23JENSEN2005, 25SCACCHI2007, 31MICHLMAYR2005, 34KHANJANI2011} where defects are reported, tracked  and discussed.  
  
\textbf{Discover that the problem exists -- } In this process activity, a developer uses as a source the requirements input on developers mailing lists, project web page, wiki page, release roadmap, newsgroup, bug reporting or tracking systems in order to find an open issue that he/she could contribute to \cite{1MOCKUS2000, 4EZEALA2008, 17JENSEN2007,18TRONG2005, 19MOCKUS2002, 22YAMAUCHI2000, 25SCACCHI2007, 26SENYARD2004, 29DILLON2009}.
 
\textbf{Select or assign a problem to work on it -- } OSS developers usually self-select projects and tasks (e.g.\ bug or issues) that they will work on \cite{2SCACCHI2006, 4EZEALA2008}. In addition, OSS developers are also end users, so they tend to work on issues reported by themselves or features that they developed. In other words, they tend to work on code areas where they are most familiar \cite{1MOCKUS2000, 19MOCKUS2002}. In some projects, there is ``code ownership'', which means part of the code is created or maintained consistently by a specific(s) developer(s). This fact does not prevent another developer from contributing to this part of the project, but there is some respect in the community for these ``senior'' contributors with experience in this project area \cite{1MOCKUS2000}. However, in other OSS projects there is no ``code ownership''. The developers are just required to ask for code review from developers who maintained this part of the code before committing it \cite{18TRONG2005}.  

In OSS development there is a ``skill meritocracy'': the core developers play a leadership role where the most skillful developers respond to them. Consequently, other developers who contribute to the project respect their decisions about the project direction and activities lead the project more cohesive centralized \cite{4EZEALA2008, 8NEULINGER2016}. Moreover, less skillful developers do not have the possibility to change the code base; they can develop or test the solution and send it to the corresponding assigned committer \cite{18TRONG2005, 8NEULINGER2016}. Notwithstanding, in the majority of the cases the developers select and assign their own tasks. Hence, on the FreeBSD project, a developer (committer) can assign a pull request to another developer who should be able to solve the problem \cite{18TRONG2005}. Regarding ``OSS 2.0'', a movement to leverage the OSS community for business, paid developers are assigned to work in OSS projects in order to find a solution to project's open problems \cite{30FITZGERALD2006}.

\textbf{Identify a solution -- } Once a developer is assigned to work on a problem, the next step is to try to identify a solution for it. The challenge most faced at this stage is to decide among the several solutions which one is better. In some cases, when a user reported a bug, for instance, he/she proposed a solution together \cite{18TRONG2005}; other developers could also have suggested alternative solutions  \cite{1MOCKUS2000, 19MOCKUS2002}. When many solutions are available, the developer forwards the solutions to the community (e.g.\ mail list or Bugzilla), aiming to get feedback from the other project group members before deciding to develop a solution \cite{1MOCKUS2000, 19MOCKUS2002}. 

\textbf{Develop the code  -- } This process activity consists basically in writing code for a solution \cite{30FITZGERALD2006} which is an essential activity in any software development process. Therefore, almost all included papers mention this activity \cite{1MOCKUS2000, 2SCACCHI2006, 4EZEALA2008, 5ABDOU2012,7KUMAR2014, 11SCOTTO2007, 13WANG2015, 14HASHMI2010, 17JENSEN2007, 18TRONG2005, 19MOCKUS2002, 22YAMAUCHI2000, 23JENSEN2005, 25SCACCHI2007, 26SENYARD2004, 27POTDAR2004, 28FUGGETTA2003, 29DILLON2009, 30FITZGERALD2006, 31MICHLMAYR2005, 32SCHROEDER2004, 33FITZGERALD2005}. According to \cite{1MOCKUS2000}, driven by passion, OSS developers write code with more care and creativity.  OSS projects have coding guidelines consisting in coding  best practices to assist the coding activity \cite{26SENYARD2004, 24SCACCHI2004}. 

\textbf{Commit the code -- } Commit the code consists of adding source code changes to the project repository in order to make it available to the community \cite{1MOCKUS2000, 18TRONG2005, 17JENSEN2007}. The authority to commit code is different in each project. As stated by \cite{14HASHMI2010}, the big part of OSS projects allow core developers (most experienced developers) to commit (more than 90\%); over 60\% of the projects allow commits from developers who contribute frequently in the project and just over 25\% of the projects allow passive developers to commit \cite{8NEULINGER2016}. In the FreeBSD project \cite{18TRONG2005}, a committer is a developer who has the authority to commit changes in the project's CVS repository.  A developer can assume the role of committer after actively contributing to the project for the last 18 months, or if the core developer team gives him/her this privilege; however this developer will have a mentor to supervise his/her work until he/she has gained experience and becomes reliable \cite{18TRONG2005}. It is important to highlight that the committer is not necessarily the code author \cite{8NEULINGER2016}: a committer could be just the developer responsible for committing changes officially in the code repository \cite{34KHANJANI2011, 14HASHMI2010, 13WANG2015, 19MOCKUS2002}.

\textbf{Testing -- } Testing is an essential activity in any software development process. OSS community has the advantage of great user involvement and a structured approach to handle bugs \cite{14HASHMI2010}.  These points collaborate with test activities leading to produce a high-quality software product \cite{20ABDOU2015}. The huge majority of the included papers mentioned software testing activity \cite{1MOCKUS2000,4EZEALA2008, 5ABDOU2012, 7KUMAR2014, 10KRISHNAMURTHY2013, 13WANG2015, 14HASHMI2010, 17JENSEN2007, 18TRONG2005, 19MOCKUS2002, 20ABDOU2015, 22YAMAUCHI2000, 23JENSEN2005, 25SCACCHI2007, 26SENYARD2004, 27POTDAR2004, 28FUGGETTA2003, 29DILLON2009, 30FITZGERALD2006,31MICHLMAYR2005, 32SCHROEDER2004, 33FITZGERALD2005, 34KHANJANI2011}.  Unit test is the most common test approach in OSS development \cite{14HASHMI2010} followed by pre-release testing \cite{14HASHMI2010, 19MOCKUS2002, 18TRONG2005}. Regarding multiple test techniques, a survey conducted by \cite{14HASHMI2010}, showed that the majority of the respondents mentioned the adoption of functional testing including some form of system testing, regression test, integration test or acceptance testing. On the other hand, \cite{26SENYARD2004} affirm that regression test is not performed in OSS projects or is not mandatory \cite{14HASHMI2010, 19MOCKUS2002}. 

The study performed by \cite{5ABDOU2012} outlined four OSS testing activities. In the following, they are described in their order of execution: (i)~Identify a test to be done --  There is no prioritization of the features that need to be tested, the features to be tested are selected independently based on developers' interests; 
(ii)~Derive testable aspects and test cases -- The testable aspects are simply based on specific attributes of interest from developers who decide to test the feature. The test cases are derived by determining pre-conditions, selecting input values and defining acceptance criteria. Moreover, approximately half of the projects analyzed by \cite{14HASHMI2010} use a type of documented test cases; 
(iii)~Execute all test cases -- This activity is performed through the test execution in a local source code copy \cite{19MOCKUS2002}. Considering that it is a manual activity, it depends on the human factors (developer expertise and judgment) \cite{19MOCKUS2002}; 
(iv)~Review or vote until accepting the test -- After executing the test, the developer commits the test results or posts the results on a bug report system (e.g.\ Bugzilla) or mailing lists. 

The second more acknowledged form of test in an OSS environment is a pre-release test (also known as system test). It consists of introducing a release candidate to the OSS project community and let them test and make fixes until a core developer team decides that the system is ready for a final release \cite{18TRONG2005}. It is important to mention that OSS users act as beta testers in OSS projects \cite{27POTDAR2004}.

\textbf{Code review -- } Code review is one of the most important peer review activities in the OSS development process. Its most important element is code inspection by other developers \cite{14HASHMI2010}. The code review practice is mentioned in the studies \cite{13WANG2015, 14HASHMI2010, 17JENSEN2007, 18TRONG2005, 19MOCKUS2002, 24SCACCHI2004, 25SCACCHI2007}.
According to \cite{14HASHMI2010}, code review is performed asynchronously and distributively and it is done before and after a source code is to be committed in the project repository. However, in \cite{18TRONG2005} the developer is required to ask for a code review before committing in a code portion that is maintained by other committers.

As reported by \cite{13WANG2015}, the code review activity can be divided in there steps: (i)~pre-review, which is performed by developer's code author and a committer; (ii)~review, which is performed by a developer acting as a code reviewer, verifier and approver; and (iii)~post-review, performed by a submitter. Not just core developers perform code review, and other developers can assume this role. However, the post-review step is usually performed by core developers since they have the power to perform project releases.

\textbf{Continuous Integration (CI) -- }  Continuous integration was addressed by a single included study. This activity is a process that integrates code changes automatically in a shared repository responsible for compiling, building and executing test cases every time a code change is submitted \cite{3RHAMAN2018}. In this study, the authors conclude that continuous integration is beneficial for OSS since after its adoption, the collaboration in larger OSS projects as well as the bug and issues resolution increased. A typical continuous integration process has  the following steps: (i)~A developer commits his/her code changes in a repository maintained by a version control system (e.g.\ Github); (ii)~These commits trigger CI jobs on a CI tool which executes, builds and tests the code; (iii)~this step results into a build feedback that is provided to developers (via e-mail and/or cellphone alerts). Based on this feedback, developers can make code changes if needed and then repeat the process until no more changes be necessary and then moving forward to the next OSS process activities.

\textbf{Perform a partial release or pre-release -- } Some OSS projects perform a pre-release before the final release in order to make sure that the project is ready to be released. Pre-releases are usually expressed as tentative alpha, beta, candidate or stable releases \cite{24SCACCHI2004, 25SCACCHI2007}. The main objective of a pre-release is to get user feedback  \cite{14HASHMI2010, 19MOCKUS2002}. According to \cite{27POTDAR2004}, users are the best beta testers. When an issue is detected, it is reported to the project community and a new development cycle is started. A pre-release is tested and fixed until the core development team decides that the system is ready for a final release \cite{18TRONG2005}. 

\textbf{Document user instructions -- } In some OSS projects is common to have a place to present project's user documentation; for example, project webpage, wiki, release roadmap lists and newsgroups \cite{1MOCKUS2000,8NEULINGER2016, 10KRISHNAMURTHY2013, 14HASHMI2010, 15SCACCHI2008, 18TRONG2005, 22YAMAUCHI2000, 23JENSEN2005, 24SCACCHI2004, 25SCACCHI2007}. Before a final release is made available, these pages must be updated, reporting the project changes and feature modifications \cite{10KRISHNAMURTHY2013}. 

\textbf{Freeze the project -- } In order to release a real stable project version, the project is frozen \cite{19MOCKUS2002, 18TRONG2005, 8NEULINGER2016} which means that only serious bug fixes \cite{8NEULINGER2016, 19MOCKUS2002} and security repairs are allowed in this period \cite{18TRONG2005}. In other words, just a bit of polishing is allowed \cite{19MOCKUS2002}. During the freeze period, widespread testing is performed until the release gets ready \cite{18TRONG2005, 19MOCKUS2002}. The freeze period is typically just a few days (e.g.\ 15 days) \cite{19MOCKUS2002, 18TRONG2005}. 

\textbf{Perform product/feature release -- } Release management is a crucial part in OSS projects \cite{14HASHMI2010}. This activity is extremely important since it is the moment to make developers collaboration results available to the OSS community. All releases are performed in a coordinated way. OSS projects have a central person or a team that controlled the official project releases \cite{1MOCKUS2000, 14HASHMI2010, 17JENSEN2007}. When it is close to a release point, a member of the core development team (most experienced project developers) assumes the role of release manager \cite{1MOCKUS2000, 18TRONG2005, 14HASHMI2010}. The survey conducted by \cite{14HASHMI2010} showed that more than half of OSS project leaders have full release authority and under 20\% give to the core development team the authority to release. On the other hand, it is clear that the consensus of the core development team is the basis for a release. 

Regarding the release time-frame, the results presented in \cite{14HASHMI2010} study demonstrated that one-third of the OSS projects considered in their study release every six months and approximately 10\% every quarter. However, the study concludes that the release frequency is ultimately up to core developers decision.
After a code release, only small code updates addressing specific problems in the last release are allowed (hotfixes) \cite{8NEULINGER2016}.

\subsubsection{OSS process roles}
\label{processroles}

In summary, the roles in OSS development can be divided into two major group visions: users and developers \cite{13WANG2015}. However, it is known that OSS developers are users too \cite{4EZEALA2008, 25SCACCHI2007}. The onion model and the layered community structure are the best-known examples of role organization in OSS projects. They are similar and present the two major group visions: (i)~developers -- core developers and others developers who contribute actively in the project (such as active developers and co-developers) or occasionally (such as peripheral developers); (ii) users -- active users who report bugs and passive ones who only use the software \cite{13WANG2015}. In this research, we identified and categorized the roles mentioned in our set of included studies. The roles identified are:

\textbf{Core developer -- } Core developers are the most experienced developers responsible for controlling and managing an OSS product \cite{34KHANJANI2011}. In other words, they control and guide the project development direction, the project architecture and perform release management \cite{1MOCKUS2000, 4EZEALA2008, 2SCACCHI2006, 5ABDOU2012, 18TRONG2005}. In the majority of OSS projects, core developers are the project founders\cite{4EZEALA2008}. According to the repository mining analysis performed by \cite{11SCOTTO2007}, core developers represent 20\% of the total developers and they develop approximately 80\% of the project's code \cite{5ABDOU2012}.

\textbf{Release manager --}  A release manager is a central person or body responsible for selecting a subset of code and perform a code release. The release manager is one of the most experienced core developers who volunteers himself/herself to assume this role when a project release needs to be done. On each release, a different core developer can assume this role \cite{1MOCKUS2000, 14HASHMI2010, 18TRONG2005}. 

\textbf{Contributor or active developer -- } A contributor or an active developer is responsible for submitting patches \cite{5ABDOU2012, 13WANG2015, 25SCACCHI2007} and in some cases, review or revise code developed by other developers \cite{13WANG2015}. They have a direct impact on the project software development since their contributions to the code base are essential to the project evolution \cite{34KHANJANI2011}. More than half of the projects surveyed by  \cite{14HASHMI2010} showed that active developers have the ability to commit code. The core development team can give the privilege to commit to an active developer \cite{16DTRONG2004}.  However, their submissions are reviewed or checked by a core developer \cite{16DTRONG2004, 18TRONG2005}. 

\textbf{Active users -- } Active users are OSS practitioners who test, report bugs and use cases to the project but they do not write code \cite{5ABDOU2012, 13WANG2015, 18TRONG2005}.

\textbf{Passive users -- } A passive user only uses the OSS software \cite{13WANG2015} and/or follows the project to stay informed about its development \cite{5ABDOU2012}. As mentioned by \cite{14HASHMI2010} and detected during the conduction of this study, not all types of roles presented in this study existing in all OSS projects also their names may vary. 
It is important to mention that the role of a volunteer in an OSS project may change over time \cite{4EZEALA2008}. For example, a very participative committed contributor can become a core developer.

The huge majority of developers in an OSS environment are volunteers. However, in some cases paid developers are assigned to work on OSS projects when a company is promoting or participating in an OSS project \cite{30FITZGERALD2006, 11SCOTTO2007}.


\section{Discussion}
\label{DiscussionandRecommendations}


In the last years, OSS processes have not been generally investigated. However, specific OSS activities or practices have been explored by researchers. This fact shows that OSS researchers are more interested in investigating OSS activities individually to understand these activities and possibly apply them in other OSS projects or even in SE traditional projects, if possible; and to improve these activities in OSS development environment.

It is recognized by OSS community the need of understanding OSS process \cite{14HASHMI2010, 2SCACCHI2006}. OSS processes followed by OSS projects were not mapped or even well documented. The existing documentation in OSS can be divided in two categories: (i) user documentation that are user instructions documented in wiki, project webpages, newsgroups and release roadmap lists; and (ii) developer documentation that are messages from mailing lists, bugs reports, code guidelines, etc. \cite{31MICHLMAYR2005, 14HASHMI2010}. The present study provides a generalized mapped process documented using BPMN notation and a detailed process activities description.

According to \cite{12LAVAZZA2010} OSS is characterized by an unstructured working environment. However, our study shows OSS projects have an organizational sense. Most of the OSS projects started with a ``seed" code \cite{4EZEALA2008, 26SENYARD2004}. In other words, the majority of OSS projects started with the characteristic of a cathedral development style \cite{26SENYARD2004}. In contrary to what is stated in the "Bazaar" model \cite{Raymond1999} which is an OSS model connoted as chaotic but effective, large OSS projects  have a systematic organization sense and care about process. For example, before the development of Apache Server project started the main developers tried to solve process issues first. They were aware that an unique development process to make decisions was needed to organize work performed by developers spread around the world without any organizational tie \cite{1MOCKUS2000}. Besides,  modification in OSS projects is not performed in an ordered way, developers focus on a modification per time (they do not mix different kinds of modifications). It is worth highlighting that successful OSS projects have governance rules that developers must follow. For example: restricted access to the source code repository; permission to commit to the project’s shared repository for release and redistribution (they need to send modifications to core developers that will evaluate and decide if they accept or not the contributions) \cite{11SCOTTO2007, 25SCACCHI2007}; and guidelines and code style standards \cite{26SENYARD2004}. 

Another observed OSS strong process characteristic is peer review activities. Peer review is the evaluation of a product work by other \cite{13WANG2015}. It's  activities include activities such as testing, parallel debugging and code review. These activities have been performed distributed and asynchronously by millions of developers in several OSS projects \cite{14HASHMI2010, 26SENYARD2004}. According to \cite{13WANG2015} peer review is a huge advantage from community involvement in OSS projects because they have ``many eyeballs" (Eric Raymond's ``Linus Law'' \cite{Raymond1999}) looking for problems, resulting in bugs found and fixed rapidly.

What motivates developers to work in OSS project is passion \cite{12LAVAZZA2010,1MOCKUS2000}. In general, OSS project developers are end-users too \cite{25SCACCHI2007} and they are driven by solving a problem motivated by the desire to solve their own problems or simply by the prestige of being recognized for their contribution for OSS community as a good programmer \cite{27POTDAR2004}. As already mentioned, roles in OSS process are guided by meritocracy. The organization's roles are crucial to organize and support project contribution management, for example, core developers act as quality assurance members in order to maintain project architecture as well as product owners guiding the product development.

Regarding testing process activities, automated testing is just addressed by one included study \cite{31MICHLMAYR2005} that concludes OSS projects spend little time effort on implementing automated tests. The authors attribute the reason for it to the OSS development nature which consists of a community that provides bug reports frequently. However, this study was published almost 15 years ago, so other investigation criteria can be applied to study automated tests specifically.

Several studies addressed by our set of included studies report process information about extremely successful OSS projects such as Apache Server \cite{1MOCKUS2000, 17JENSEN2007, 5ABDOU2012, 13WANG2015, 19MOCKUS2002}, FreeBSD \cite{18TRONG2005}, Mozilla Firefox \cite{19MOCKUS2002, 17JENSEN2007, 5ABDOU2012, 13WANG2015}, Moodle \cite{10KRISHNAMURTHY2013}, ILIAS \cite{10KRISHNAMURTHY2013} and NetBeans IDE \cite{23JENSEN2005, 17JENSEN2007, 15SCACCHI2008,13WANG2015}. Moreover, we systematically analyzed the state-of-art of existing research on OSS process. Based on that, we conclude macro process and process activities characteristics described in this work generalized OSS process. Nevertheless, it is important to make clear that every OSS project has its process particularities. 
 
Contrarily, a large number of OSS projects are stagnant \cite{25SCACCHI2007}. This main failure reason is lack of interest from OSS community driven by low project usefulness and attractiveness \cite{4EZEALA2008, 25SCACCHI2007}. As reported by \cite{4EZEALA2008}, detailed documentation is fundamental to attract contributors to collaborate in OSS projects. The absence of formal project activities and different steps performed during the project execution is a barrier to incentive collaboration in OSS process \cite{26SENYARD2004}. The process map and its activities characteristics description presented in our study are process documentation that can contribute to improve the understanding of OSS process activities and consequently help to mitigate OSS project failure.

\section{Recommendations}
\label{Recommendations}


This paper provides a generalized OSS process based on literature evidence. The synthesis from the included primary sources leads us to provide some recommendations to the OSS community. They are described following.

Our study shows that there is a process with activities and roles well defined behind OSS project success. Also, it is clear that an organized process facilitates project development and improves collaboration and engagement. Therefore, we recommend the OSS community to clearly document and keep up-to-date their process  on their documentation page or wiki, in order to clarify for everyone what are the process activities and roles. Diagrams such as BPMN are an interesting form of presenting these process activities since they are easy to read and understand.

Another core point in the OSS environment is peer review activities. Testing,  parallel debugging and code review are activities that need to be systematically integrated into an OSS process. Continuous integration is an activity with the potential to facilitate not just the integration of the code changes from multiple contributors but improve peer review practices. For example, CI best practices like Test-Driven Development (TDD) (an activity that needs to be further explored in the OSS context) and pull requests with code reviews can increase the code assertion, reduce technical debt and consequently increase delivery quality. In this context, we strongly advise considering CI practices in OSS projects.

Last but not least, project communication channels such as mails lists, bug tracking systems, web pages, wikis, etc.\ are the primary sources of OSS requirements. Best practices for their use can facilitate problem-solving and requirements elicitation.

\section{Threats to Validity}
\label{threatstovalidity}
We report on the main threats to validity of our study as well as the adopted mitigation strategies. 

\textbf{Construct validity -- } This threat is related to the construction and execution of the SLR study protocol. In order to mitigate this limitation, our SLR protocol followed the guidelines of  \cite{Kitchenham07} and \cite{Wohlin14}.  We rigorously selected relevant studies following defined selection criteria. Also, we performed a search strategy pilot test as recommended by \cite{Kitchenham07} and adopted a study control group to calibrate our search string. During the search strategy execution and data extraction analysis, we performed weekly meetings among all authors whose work progress was reviewed and any disagreement on results was discussed.  

\textbf{External validity -- } The most potent external threat to the validity of our study is that our set of included studies is not large enough to represent the state-of-art of the OSS process.  In order to mitigate this threat, our search strategy included an automated search on the most renowned SE digital libraries complemented with a snowballing forward approach. Besides, our selection criteria considered the inclusion of the only peer-reviewed studies and focused exclusively on publications that addressed the OSS process in some way discarding hybrid development process data. We are aware that some specific OSS process evidence could be missed by applying our selection criteria. Thus, we focused only on papers that address the OSS process or models explicitly. As the main goal of this study is to analyze the OSS process accurately, summarize results from explicit evidence is crucial to avoid bias in our results.

\textbf{Conclusion validity -- } This threat addresses the conclusions made for building of the presented OSS macro process. The OSS macro process was built considering extracted evidence from the included primary studies. To mitigate this threat, we performed weekly meetings to discuss among the authors the findings extracted from the 33 included papers. During these meetings, we reviewed the OSS process and made improvements on it continuously. A final review meeting was performed in order to produce and review the final version of the OSS macro process. 

\section{Conclusions}
\label{Conclusions}

In this study, we presented an overview of how the OSS community has been investigating the OSS process over the past few years and proposed a macro OSS development process with a detailed description of its process activities characteristics through a narrative description. Our study extracted and systematically analyzed OSS studies and translated these findings using BPMN notation. 

In summary, our results showed that the OSS process has not been generally investigated by OSS researchers in the last years and that OSS processes follow a general process. Also the presented OSS macro process can be used as a guide for OSS projects and also being adapted according to the OSS project reality as well as provide insights to managers and developers who want to improve their development process even in OSS and traditional environments. Finally, we presented recommendations for OSS community regarding OSS process activities.


As future work, we intend to apply the findings of our study in more recent OSS projects and provide a validation of the proposed OSS macro process though a practical point of view analysing OSS projects process activities, their characteristics and understanding how roles are involved in each activity, fact that still not clear yet in the literature.
Furthermore, we intend to investigate how researchers perform OSS process analysis in OSS projects, including approaches, techniques and tools they have used to retrieve OSS process information.

\bibliographystyle{unsrt}  

\bibliography{references}  

\end{document}